\documentclass[aps,prd,12pt,nofootinbib,floatfix,preprintnumbers]{revtex4}
\usepackage{amssymb}
\usepackage{amsmath}
\usepackage[dvips]{color}
\usepackage[dvips]{graphicx}
\usepackage{epsfig}

\unitlength=1.2mm \preprint{JLAB-THY-07-687}
\begin{document}
\newcommand{\tr}{\mbox{tr}\,}
\newcommand{\Dslash}{{\mathchoice
    {\Dslsh \displaystyle}%
    {\Dslsh \textstyle}%
    {\Dslsh \scriptstyle}%
    {\Dslsh \scriptscriptstyle}}}
\newcommand{\Dslsh}[1]{\ooalign{\(\hfill#1/\hfill\)\crcr\(#1D\)}}
\newcommand{\leftvec}[1]{\vect \leftarrow #1 \,}
\newcommand{\rightvec}[1]{\vect \rightarrow #1 \:}
\renewcommand{\vec}[1]{\vect \rightarrow #1 \:}
\newcommand{\vect}[3]{{\mathchoice
    {\vecto \displaystyle \scriptstyle #1 #2 #3}%
    {\vecto \textstyle \scriptstyle #1 #2 #3}%
    {\vecto \scriptstyle \scriptscriptstyle #1 #2 #3}%
    {\vecto \scriptscriptstyle \scriptscriptstyle #1 #2 #3}}}
\newcommand{\vecto}[5]{\!\stackrel{{}_{{}_{#5#2#3}}}{#1#4}\!}
\newcommand{\vdot}{\!\cdot\!}

\bibliographystyle{apsrev}

\title{Nucleon Structure and Hyperon Form Factors from Lattice QCD}

\author{Huey-Wen Lin}
\email{hwlin@jlab.org} \affiliation{Thomas Jefferson National
Accelerator Facility, Newport News, VA 23606
\\RIKEN-BNL Research Center, Brookhaven National
Laboratory, Upton, NY 11973
}
\date{July, 2007}
\pacs{11.15.Ha,12.38.Gc,12.38.Lg,14.40.-n}
\begin{abstract}
In this work, I report the latest lattice QCD calculations of nucleon and hyperon structure from chiral fermions in 2+1-flavor dynamical simulations. All calculations are done with a chirally symmetric fermion action, domain-wall fermions, for valence quarks.
I begin with the latest lattice results on the nucleon structure, focusing on results from RBC/UKQCD using 2+1-flavor chiral fermion actions. We find the chiral-extrapolated axial coupling constant at physical pion mass point to be 1.23(5), consistent with experimental value.
The renormalization constants for the structure functions are obtained from RI/MOM-scheme non-perturbative renormalization.
We find first moments of the polarized and unpolarized nucleon structure functions at zero transfer momentum to be 0.133(13) and 0.203(23) respectively, using continuum chiral extrapolation. These are consistent with the experimental values, unlike previous calculations which have been 50\% larger. We also have a prediction for the transversity, which we find to be 0.56(4). The twist-3 matrix element is consistent with zero which agrees with the prediction of the Wandzura-Wilczek relation.

In the second half of this work, I report an indirect dynamical estimation of the strangeness proton magnetic moments using mixed actions. With the analysis of hyperon form factors and using charge symmetry, the strangeness of proton is found to be $-0.066(26)$, consistent with the Adelaide-JLab Collaboration's result. The hyperon $\Sigma$ and $\Xi$ axial coupling constants are also performed for the first time in a lattice calculation, $g_{\Sigma\Sigma}= 0.441(14)$ and $g_{\Xi\Xi} = -0.277(11)$.
\end{abstract}

\maketitle
\section{Introduction}

Quantum chromodynamics (QCD) has been successful in describing many properties of the strong interaction. In the weak-coupling regime, we can rely on perturbation theory to work out the path integral which describes physical observables of interest. However, for long distances perturbative QCD no longer converges. Lattice QCD allows us to calculate these quantities from first principles.

In Lattice QCD, space and time are discretized in a finite volume, and we use Monte Carlo integration to evaluate the remaining integral. Since the real world is continuous and infinitely large, at the end of the day we will have to take $a \rightarrow 0$ and $V \rightarrow \infty$. However, using current computer resources, we cannot yet simulate full QCD at the physical pion mass. With the help of the chiral perturbation theory and calculations at multiple heavier pion masses which are affordable in terms of available computational resources, we can extrapolate quantities of interest to the physical limit. Such calculations also help to determine the low-energy constants of the chiral effective theory.

There are a few choices of fermion action that have been commonly used in lattice QCD calculations. Each has its own pros and cons. They differ primarily by how they maintain symmetry, their calculation cost and their discretization error. The consistency of results from different fermion actions demonstrates the university of discretizations from lattice QCD.

The most expensive class of discretization are the chiral fermion actions\cite{Kaplan:1992bt,Kaplan:1992sg,Shamir:1993zy,Furman:1994ky}: domain-wall fermion (DWF) or overlap fermion. Such actions maintain the chiral symmetry of the fermions at great cost, but for this cost, we derive significant benefits. The theory is automatically $O(a)$ improved which makes it particularly well suited for spin physics and weak matrix elements. Since symmetry remains at the non-zero lattice spacing, it further simplifies the renormalization calculation on the lattice and the chiral extrapolation.

A much cheaper alternative is the (improved) staggered fermion action (asqtad)\cite{Kogut:1975ag,Orginos:1998ue,Orginos:1999cr}. The relatively fast simulation has great potential to be the first lattice calculation to reach the physical pion mass with 2+1 flavors. However, it introduces the problem of taste: each single fermion in the action contains four tastes. Although these extra tastes can be removed by the ``root trick'', mixing among different parities and tastes remains in the theory. There have been spirited debates over whether it is proper to use asqtad\cite{Creutz:2006wv,Sharpe:2006re}, but for practical purposes no evidence of anomalous results have yet been found; theoretical proof of its correctness is an ongoing effort. However, the issue of taste-breaking makes baryonic operators a nightmare to work with, regardless of any potential cost savings.

One might combine the best features of both of these actions by using a mixed action. Cheap staggered fermions can be used for the expensive sea quarks, while chiral domain-wall fermions are used in the valence sector, where they protect operators from mixing. We will discuss the employment of such an action in the final section.

In this work, we use lattice QCD techniques to pursue long-distance physics, solving non-perturbative QCD from first principles. The structure of this article is as follows:
In Sec.~\ref{sec:nucleon_struc}, I report the latest 2+1f DWF valence calculation using RBC/UKQCD gauge configurations. In this work, we concentrate on the results for the axial coupling constant, first moment of the unpolarized quark and helicity distributions, and the zeroth moment of the transversity and twist-3 matrix element $d_1$. We reproduce the experimental numbers for the first three quantities using continuum perturbation theory, predict the value of the transversity at leading order and check the Wandzura-Wilczek relation. We also compare our results with those of other lattice groups.
In the Sec.~\ref{sec:hyperons}, I present some work done by the Jefferson Lab hyperon project, calculating the proton strange magnetic moment and the axial couplings of the hyperons. This is the first time that the strange magnetic moment is done using dynamical lattice data, and we find a value consistent with the Adelaide-JLab Collaboration's result. The $\Sigma$ and $\Xi$ axial coupling is for the first time done in lattice QCD. We find the numbers are more accurate than what had been estimated by the chiral perturbation theory or large-$N_c$ theory.

\section{Nucleon Structure}\label{sec:nucleon_struc}

\subsection{Lattice Parameters}\label{subsec:parameters}
In this calculation, we uses lattices generated by the RBC/UKQCD collaborations with the chiral DWF action and a full 2+1-flavor dynamical quark sector. The ensembles range in pion mass from 625 down to 300~MeV, at a lattice scale 1.6~GeV in a 3~fm box. The details of the gauge configurations can be found in Ref.~\cite{Allton:2007hx} for hadron properties in a 2~fm box.

On these ensembles, we use a Gaussian-smeared source to improve the signals. The source-sink separation is fixed at 12 time units. The choices of sink and the corresponding number of configurations in this work are listed in Table~\ref{tab:SrcPar}.

\begin{table}
\caption{2+1 flavor Gaussian-smeared source parameters}\label{tab:SrcPar}
\begin{center}
\begin{tabular}{c|c|c|c|c}
\hline\hline $m_{\rm sea}$ & 0.005 & 0.01 &  0.02 & 0.03\\
\hline
$t_{\rm src}$ &  $0,32$ & $0,16,32,48$ & $0,16,32,48$ & $0,16,32,48$ \\
\# of conf. & 180 & $119$ & $49$ & $54$ \\
$m_{\pi}$~(GeV) & 0.319(3) & 0.399(3) & 0.535(3)  & 0.625(3)  \\
$m_{N}$~(GeV)   & 1.085(16) & 1.169(19)& 1.204(13) & 1.474(18)  \\
\hline\hline
\end{tabular}
\end{center}
\end{table}

The interpolating field used in our calculation is
\begin{eqnarray}
\chi^N = \sum_{\vec{x},a,b,c}
e^{i\vec{p}\cdot\vec{x}}\epsilon^{abc}\left[u_a^T(y_1,t)C\gamma_5
d_b(y_2,t)\right] u_{c,\alpha}(y_3,t) \phi(y_1-x) \phi(y_2-x)
\phi(y_3-x)
\end{eqnarray}
We calculate the nucleon two-point function (${C_{\rm 2pt}}$ of 
$\chi^N$ with smearing parameters $A$ and $B$) as
\begin{eqnarray}
\Gamma^{(2)}_{AB}(t)&=& \langle
\chi^N(t)(\chi^N)^\dagger(t_{\rm src})\rangle
\end{eqnarray}
and three-point function which is defined as
\begin{eqnarray}\label{eq:three-pt}
&&(\Gamma^{(3)}_\mu(t_{\rm src},t,t_{\rm snk}))_{AB}\nonumber \\
&&= \langle \chi_N(t_{\rm snk},\vec p_{\rm snk})\,
{\cal O} (t,\vec{q})\,
\chi^\dagger_N(t_{\rm src},\vec p_{\rm src})\rangle,
\end{eqnarray}
where ${\cal O}$ is the operator of interest. 
For more details, please refer to our earlier work in Refs.~\cite{RBC_2f,Orginos:2005uy,Sasaki:2003jh}.

\begin{figure}
\includegraphics[width=0.45\textwidth]{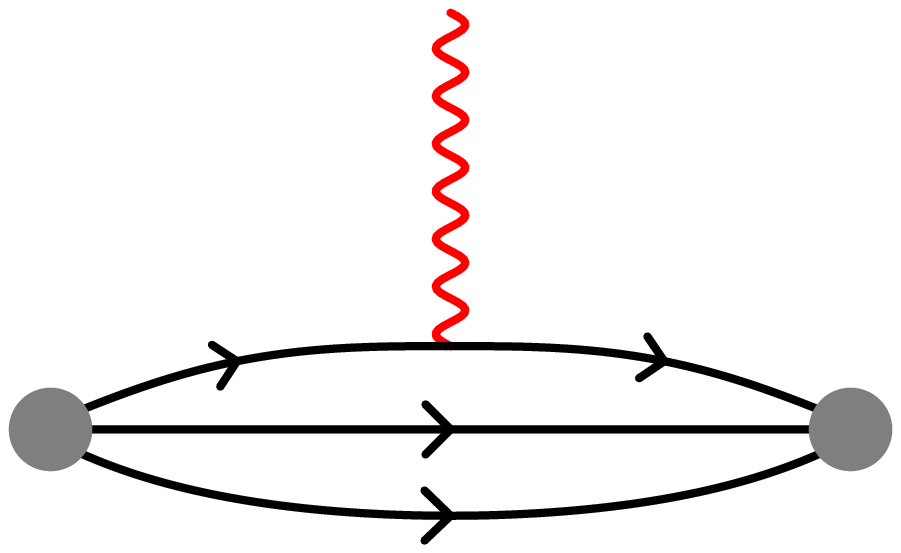}
\includegraphics[width=0.45\textwidth]{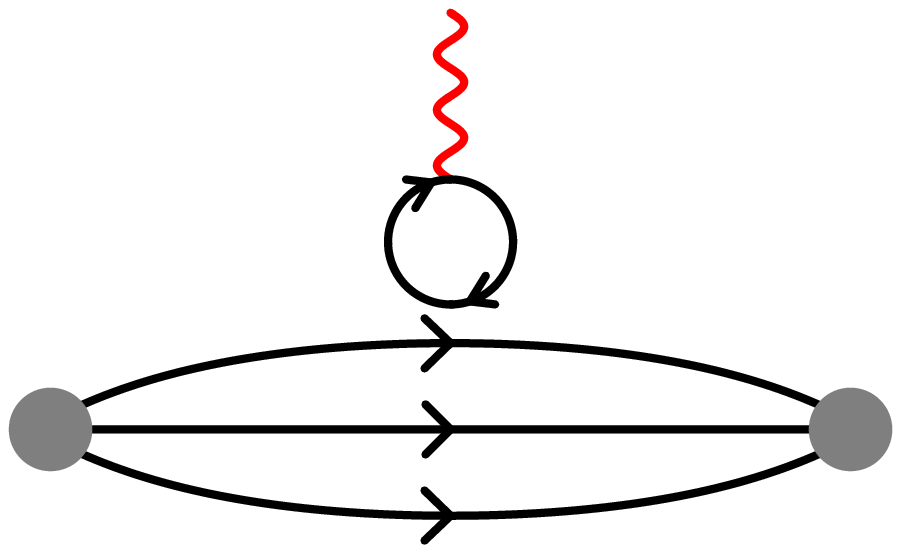}
\caption{The three-point Green function contains connected (left) and disconnected (right) diagrams. The disconnected piece is ignored in this work by focusing only on isovector quantities.}\label{fig:3pt}
\end{figure}

When one calculates the three-point Green function, there are two possible contraction topologies: ``connected'' and ``disconnected'' diagrams, as depicted in Figure~\ref{fig:3pt}. Disconnected pieces are notoriously difficult to calculate directly on the lattice\cite{Mathur:2000cf,Dong:1997xr,Lewis:2002ix,Foley:2005ac}. It would require numerous source vectors in the fermion-matrix inversion, and there was no reasonable way of calculating these contributions when this calculations started. However, this difficulty might be resolved in the near future with  development of new techniques by members of the USQCD collaboration. In this work, only ``isovector'' quantities will be discussed, in which the disconnected piece cancels under isospin symmetry.

\begin{figure}[t]
\vspace{-0.5cm}
\includegraphics[width=0.45\columnwidth]{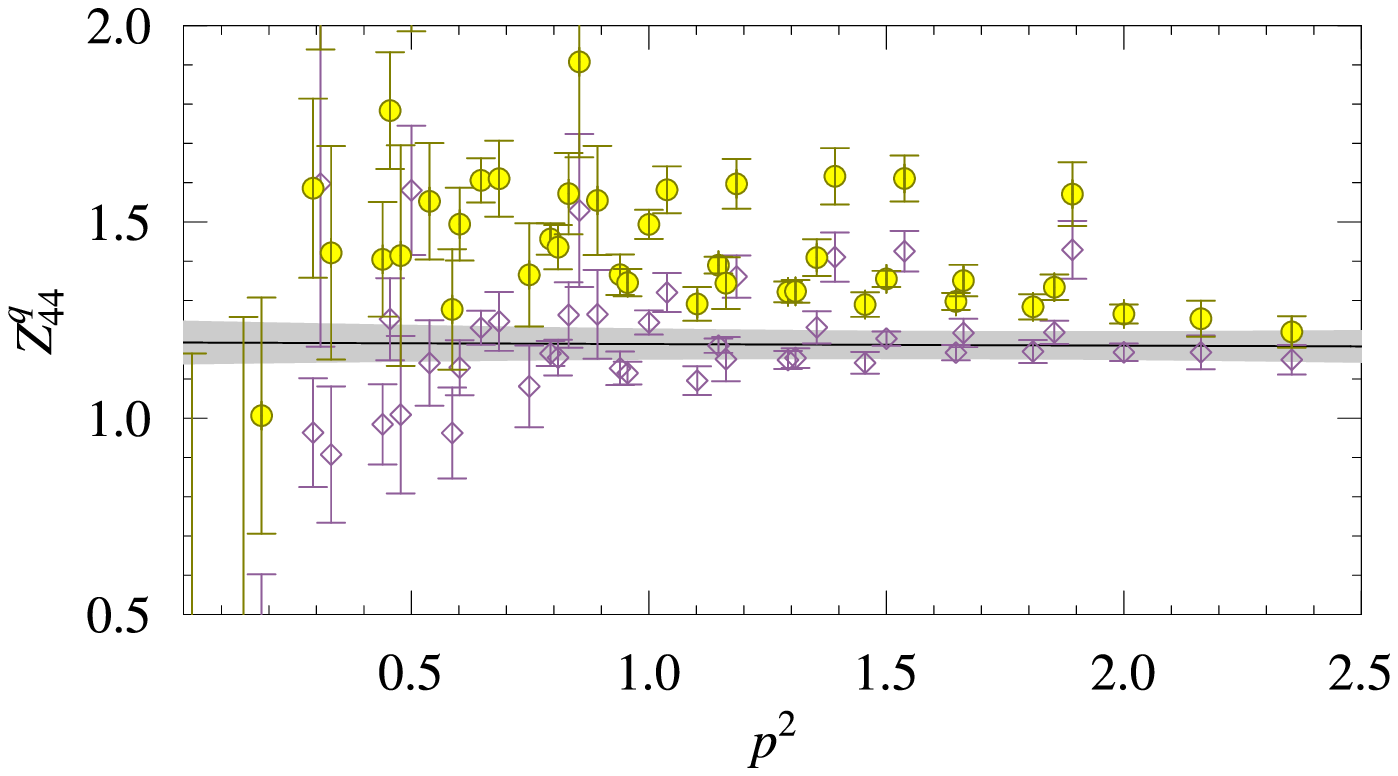}
\includegraphics[width=0.45\columnwidth]{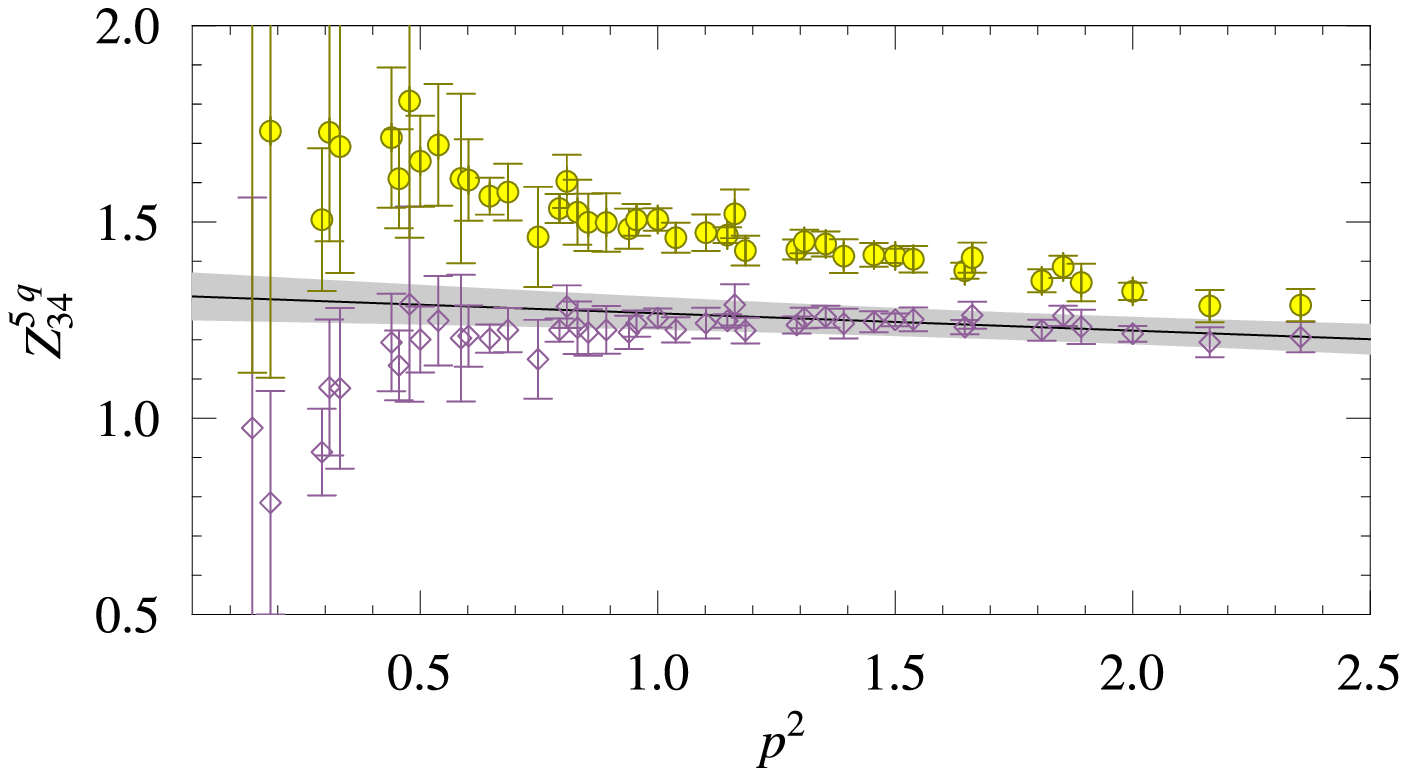}
\includegraphics[width=0.45\columnwidth]{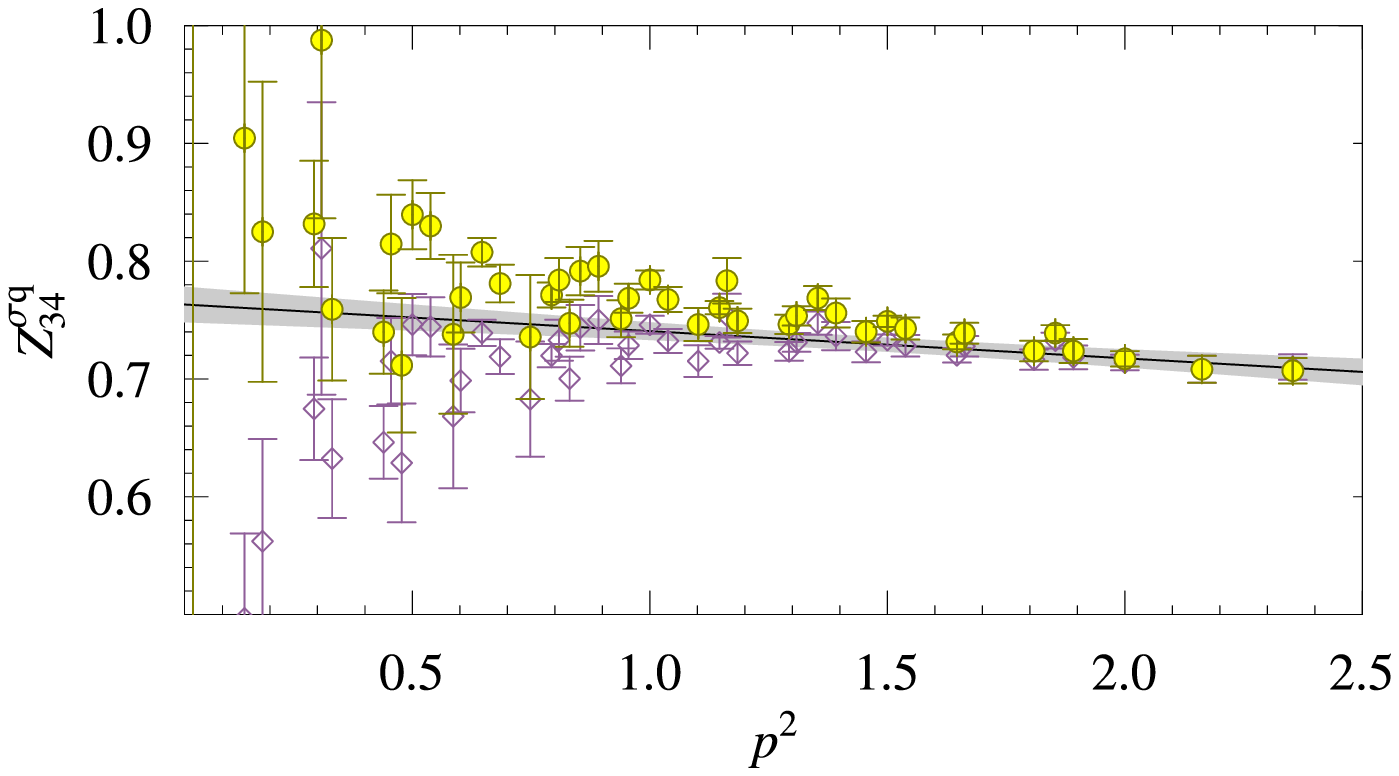}
\caption{Renormalization constants in the chiral limit. The lightly-filled circles are the renormalization constants in RI/MOM scheme, and the diamond points are $\overline{\rm MS}$-scheme at 2~GeV. The fitted lines are used to remove $(ap)^2$ artifacts.} \label{fig:npr}
\end{figure}

We apply nonperturbative renormalization (NPR) in RI/MOM scheme\cite{Martinelli:1994ty} to the above quantities. In general, the operators of interest can mix with lower-dimension operators as
\begin{eqnarray}
{\cal O}_i(\mu) = Z_i(\mu,a){\cal O}_i(\mu)+\sum_{i \neq j}
Z_{ij}(\mu;a){\cal O}_j(\mu)
\end{eqnarray}
With the good chiral symmetry of DWF action, however, we are protected from this mixing problem. We calculate $Z_{O_\Gamma}(\mu; a)$ in RI/MOM scheme, where $\mu$ must fall inside the renormalization window $\Lambda_{\rm QCD} \ll \mu \ll 1/a$. Then we convert to $\overline{\rm MS}$ scheme~\cite{Gimenez:1998ue}, running to 2~GeV to get the renormalization constant for the operators. In this work, the NPR is done on a smaller lattice ensemble, the 2~fm ones; since it is a short-distance quantity, the renormalization constants are not as sensitive to finite-volume effects as other observables. Detailed step-by-step descriptions can be found in Refs.~\cite{RBC_2f,Orginos:2005uy}. The renormalization constant in RI/MOM and $\overline{\rm MS}$ scheme for first moment of the unpolarized distribution and helicity and zeroth moment of the transversity (from top to bottom) are shown in the Figure~\ref{fig:npr}.

\subsection{Numerical Results}

\subsubsection{Axial Coupling Constant}\label{subsubsec:gA}
\begin{figure}[h]
\includegraphics[width=0.49\columnwidth]{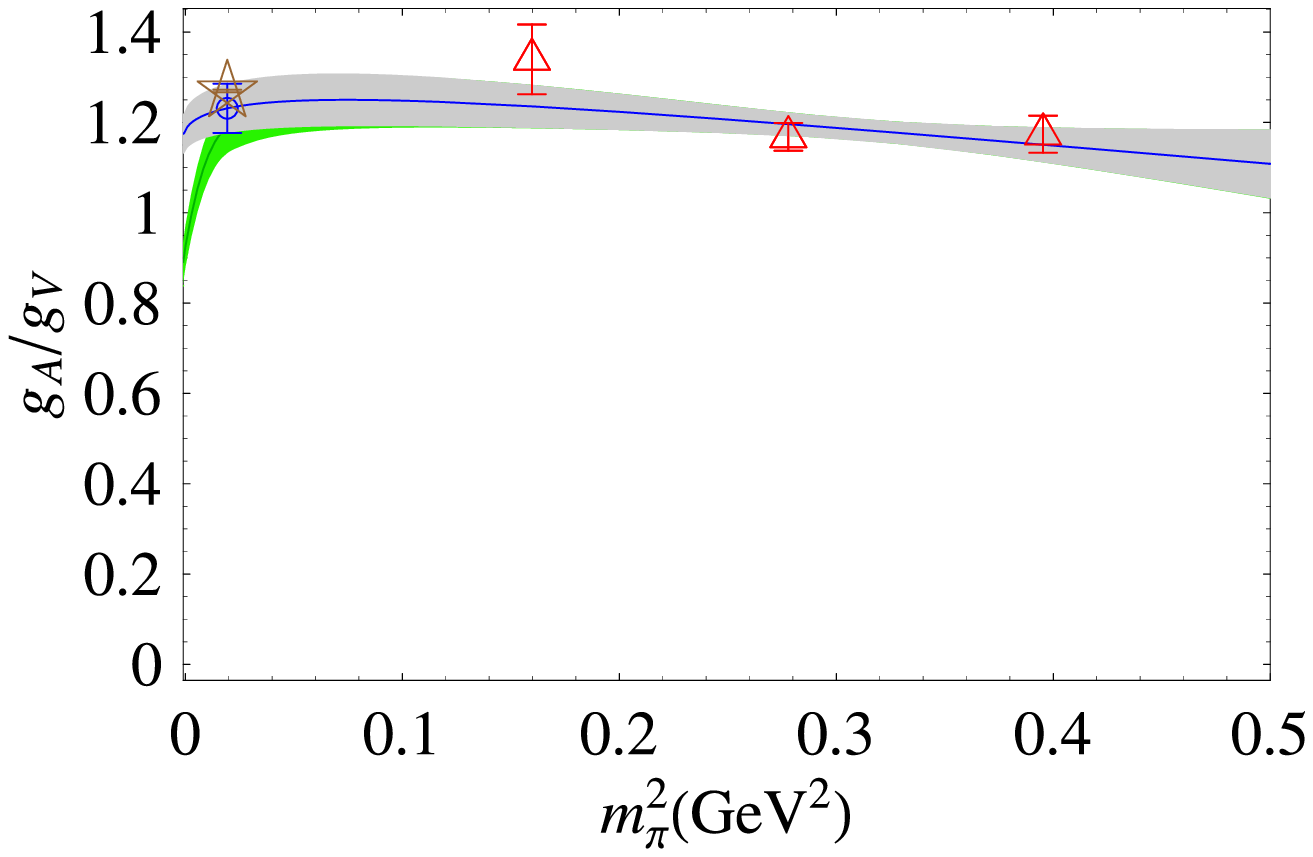}
\includegraphics[width=0.49\columnwidth]{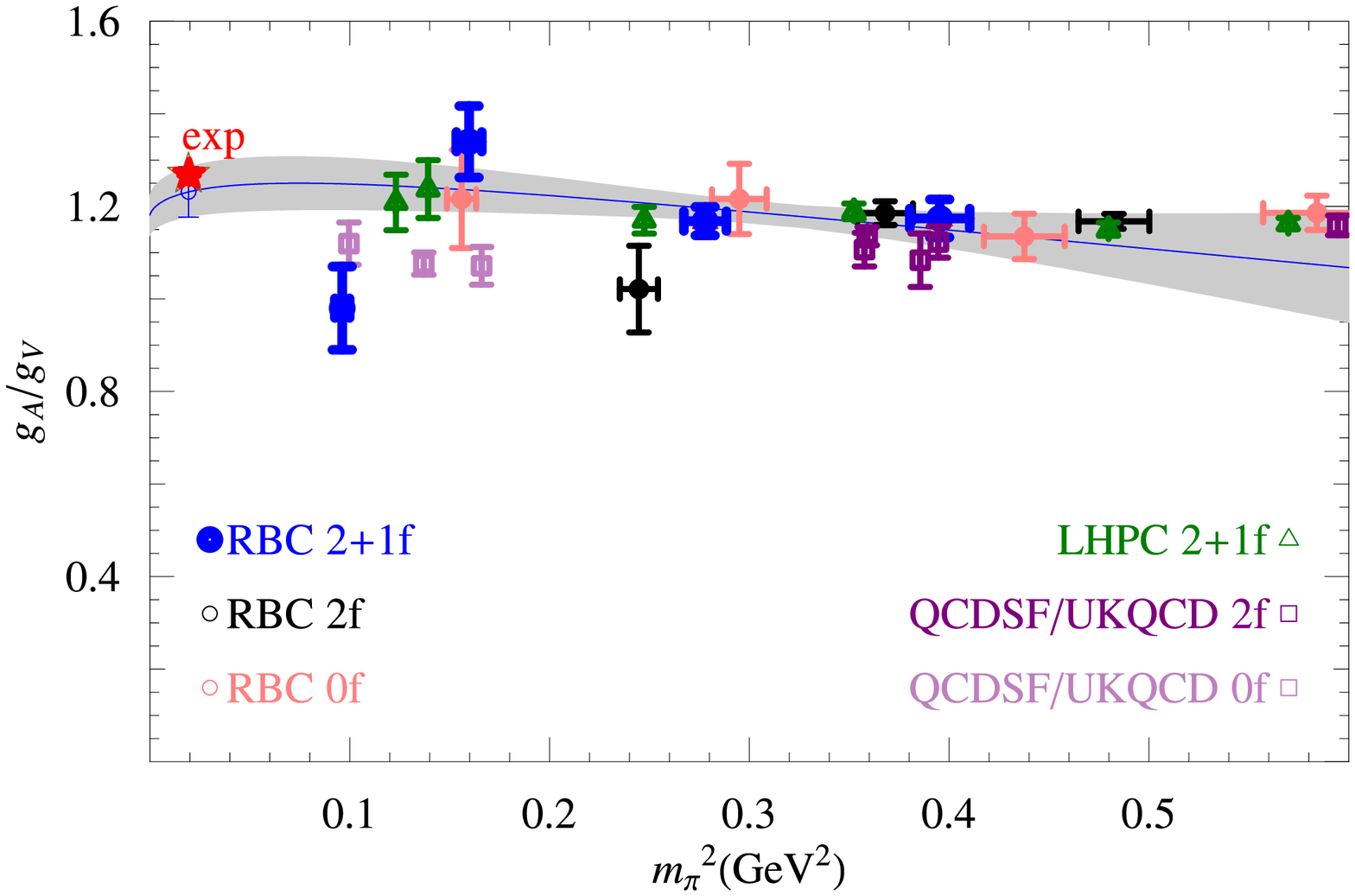}
\vspace{-0cm} \caption{(left) Renormalized axial charge in terms of
pseudoscalar mass with the small scale expansion fit  (grey band)
and an estimation of the finite-volume effect
(green band)\\
(right) The axial charge comparison from various lattice group } \label{fig:gA_all}
\end{figure}

The axial charge is well measured in the neutron $\beta$ decay experiment and hence it is a natural candidate to demonstrate how well the lattice QCD approach with the chiral extrapolation to the physical pion point.
The isovector vector and axial charges, $g_V$ and $g_A$, are defined as the zero-momentum-transfer limits of the following, 
\begin{eqnarray}
\langle p| V^+_\mu(x; q=0) | n \rangle &=& \bar{u}_p \left(\gamma_\mu
g_{V} \right){u}_n \label{eq:cont_vector} \\
\langle p| A^+_\mu(x; q=0) | n \rangle &=&
\bar{u}_p \left(\gamma_\mu  \gamma_5 g_{A} \right)  u_n . \label{eq:cont_axial}
\end{eqnarray}
Because of chiral symmetry on our fermion action, a Takahashi-Ward identity ensures that the two currents, which are related by chiral transformation, share a common renormalization: $Z_A = Z_V$ up to a lattice discretization error of $O(ma^2)$. Since the vector current is conserved, its renormalization is easily obtained as the inverse of the vector charge $g_V$. Thus, by calculating the ratio of the three-point functions for $g_A/g_V$, we get the renormalized axial charge, $(g_A)^{\rm ren}$. The results are shown as red triangles in Figure~\ref{fig:gA_all}.

In order to reach the physical pion mass value for our result, we adopt the chiral extrapolation expression from the the small-scale expansion (SSE) scheme\cite{Hemmert:1997ye}. In this scheme, one uses explicit degrees of freedom from the pion, nucleon and $\Delta(1232)$, then expands in terms of $\Delta_0$, the mass splitting between the $N$ and $\Delta$ in the chiral limit, which is treated as $O(\epsilon)$. Here we adopt a formulation which is correct up to $O(\epsilon^3)$, as seen in Ref.~\cite{Hemmert:2002uh,Hemmert:2003cb}, and we also try its finite-volume corrected form. The grey band in Figure~\ref{fig:gA_all} indicates the uncertainty due to the jackknife chiral extrapolation with SSE formulation; our preliminary extrapolated axial coupling constant is 1.23(5), consistent with experimental data.
If we add finite-volume effects, with the lattice box fixed at 3~fm, we can see the correction at our pion mass points are tiny, as indicated by the green band.
In right of Figure~\ref{fig:gA_all}, we compare our result to that of other lattice groups\cite{Edwards:2006qx,Gockeler:2006uu,Khan:2006de,Khan:2004vw,Dolgov:2002zm} and to our previous calculation~\cite{RBC_2f,Sasaki:2003jh}. Compared with our previous DWF quenched 2.4~fm and 2-flavor 2~fm box calculations, we see a clear consistency in the axial charge coupling. In fact, most of the calculations (either using chiral fermion or not) are consistent with each other. One thing to notice is that we currently have the lightest dynamical pion mass point, although the result from the lightest ensemble (in both our 2- and 2+1-flavor) deviates from the chiral extrapolation curve. We have more gauge ensembles in the 2+1-flavor case, and we will be able to verify in the near future whether this is just simply due a lack of statistics or to a finite-volume effect that is not accounted accurately by the chiral perturbation theory.

\subsubsection{Unpolarized quark and helicity distribution }\label{subsec:moments}

The moments of the unpolarized quark distribution and helicity distribution are defined as:
\begin{eqnarray}
\langle x^n\rangle_{q} &=& \int_0^1 dx x^n[q(x)-(-1)^n\overline{q}(x)]\\
\langle x^n\rangle_{\Delta q} &=& \int_0^1 dx x^n[\Delta q(x)+(-1)^n\Delta \overline{q}(x)],
\end{eqnarray}
where $q$ is the sum of the quarks with helicity aligned and anti-aligned ($q^\uparrow+q^\downarrow$) and $\Delta q$ is the difference of the two ( $q^\uparrow-q^\downarrow$).
On the lattice, this corresponds to the matrix elements of the operators
\begin{eqnarray}
{\cal O}_{\mu_1...\mu_n}^q &=& i^{n-1} \overline{\psi} \gamma^{\{\mu}\overleftrightarrow{D}_{\mu_2}\cdot \overleftrightarrow{D}_{\mu_n\}}\psi \\
{\cal O}_{\mu_1...\mu_n}^{5q} &=& i^{n-1} \overline{\psi} \gamma_{\{\mu} \gamma_5 \overleftrightarrow{D}_{\mu_2}\cdot \overleftrightarrow{D}_{\mu_n\}}\psi
\end{eqnarray}
respectively with $\overleftrightarrow{D}
=\frac{1}{2}(\overrightarrow{D}-\overleftarrow{D})$ the difference between the covariant derivatives. Note that here the trace term which corresponds to the disconnected piece is not included in our calculation. Therefore, we will look at the difference between the up and down quark contribution where this contribution is negligible.

\begin{figure}
\vspace{-0.5cm}
\includegraphics[width=0.49\columnwidth]{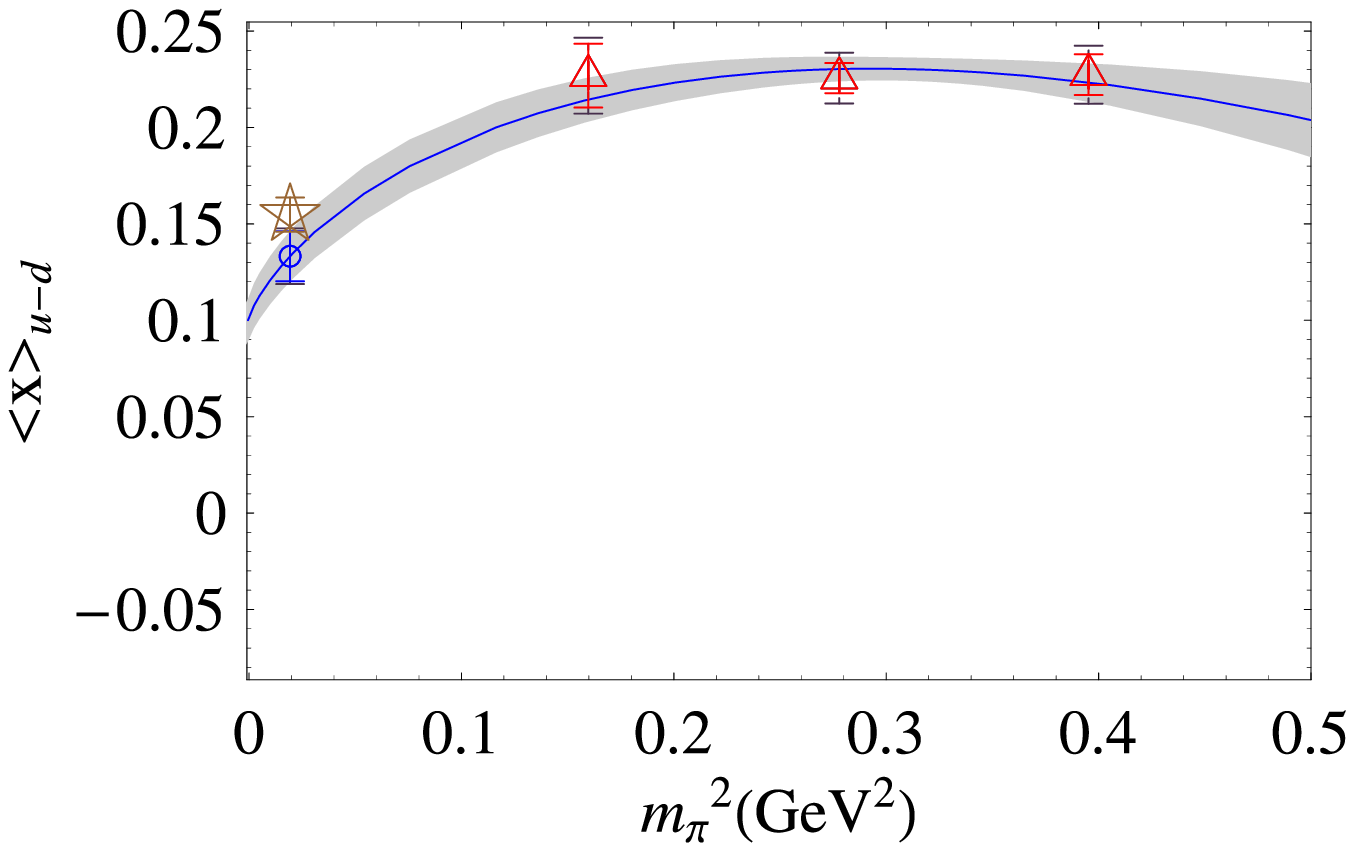}
\includegraphics[width=0.49\columnwidth]{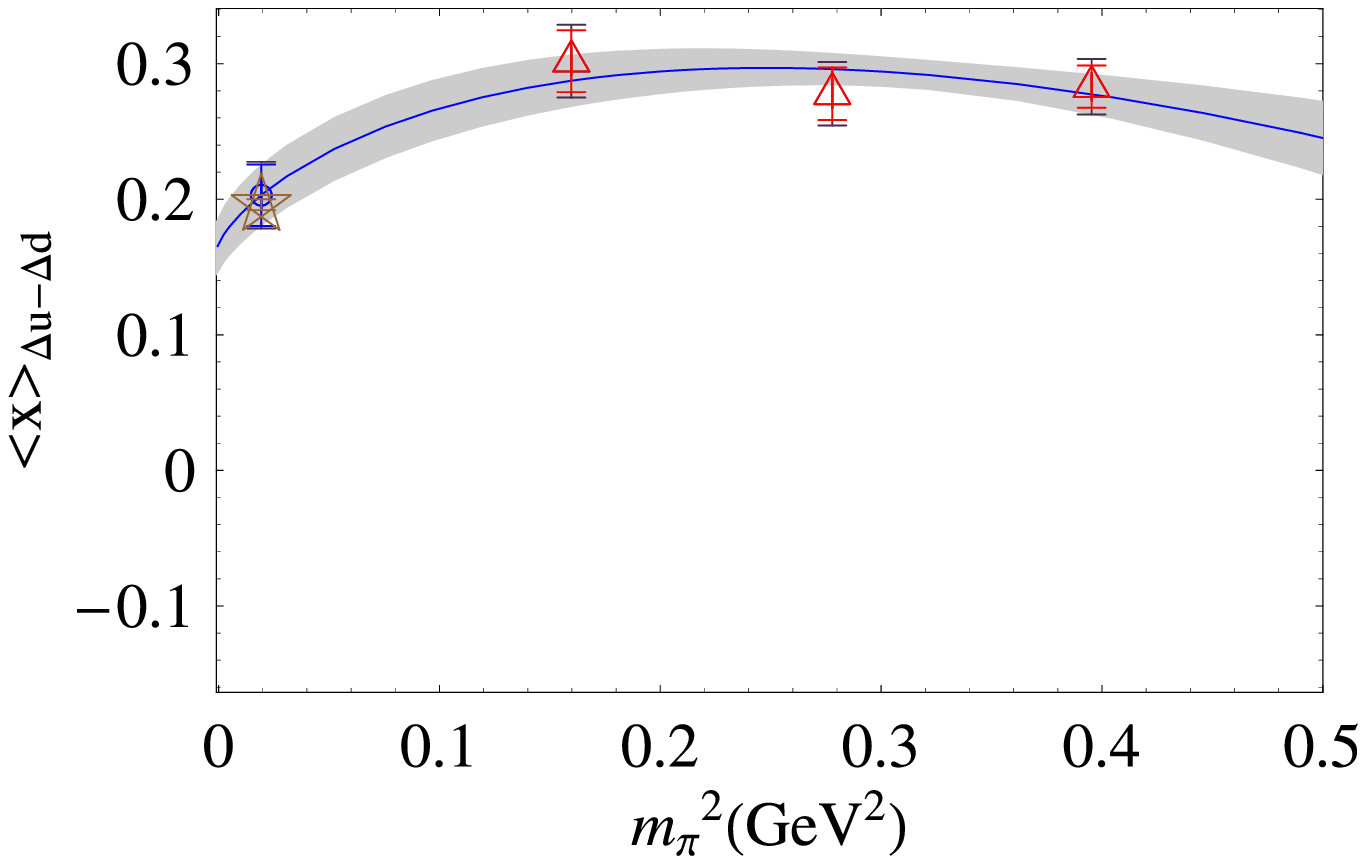}
\vspace{-0cm} \caption{Renormalized first moment of the unpolarized (left column) and helicity (right column) distributions, in terms of $m_\pi^2$ and their chiral extrapolations} \label{fig:moments}
\end{figure}

\begin{figure}
\vspace{-0.5cm}
\includegraphics[width=0.49\columnwidth]{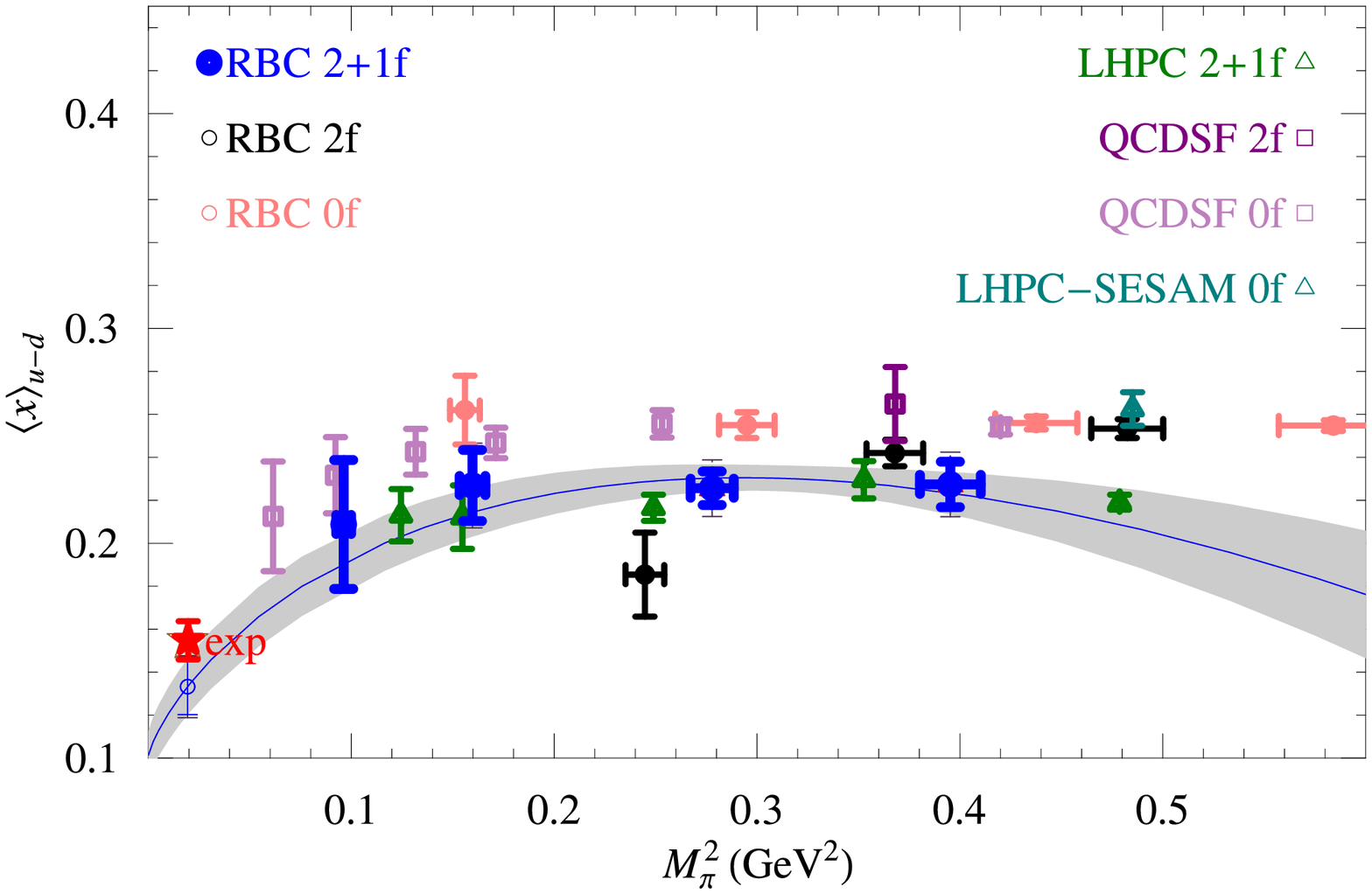}
\includegraphics[width=0.49\columnwidth]{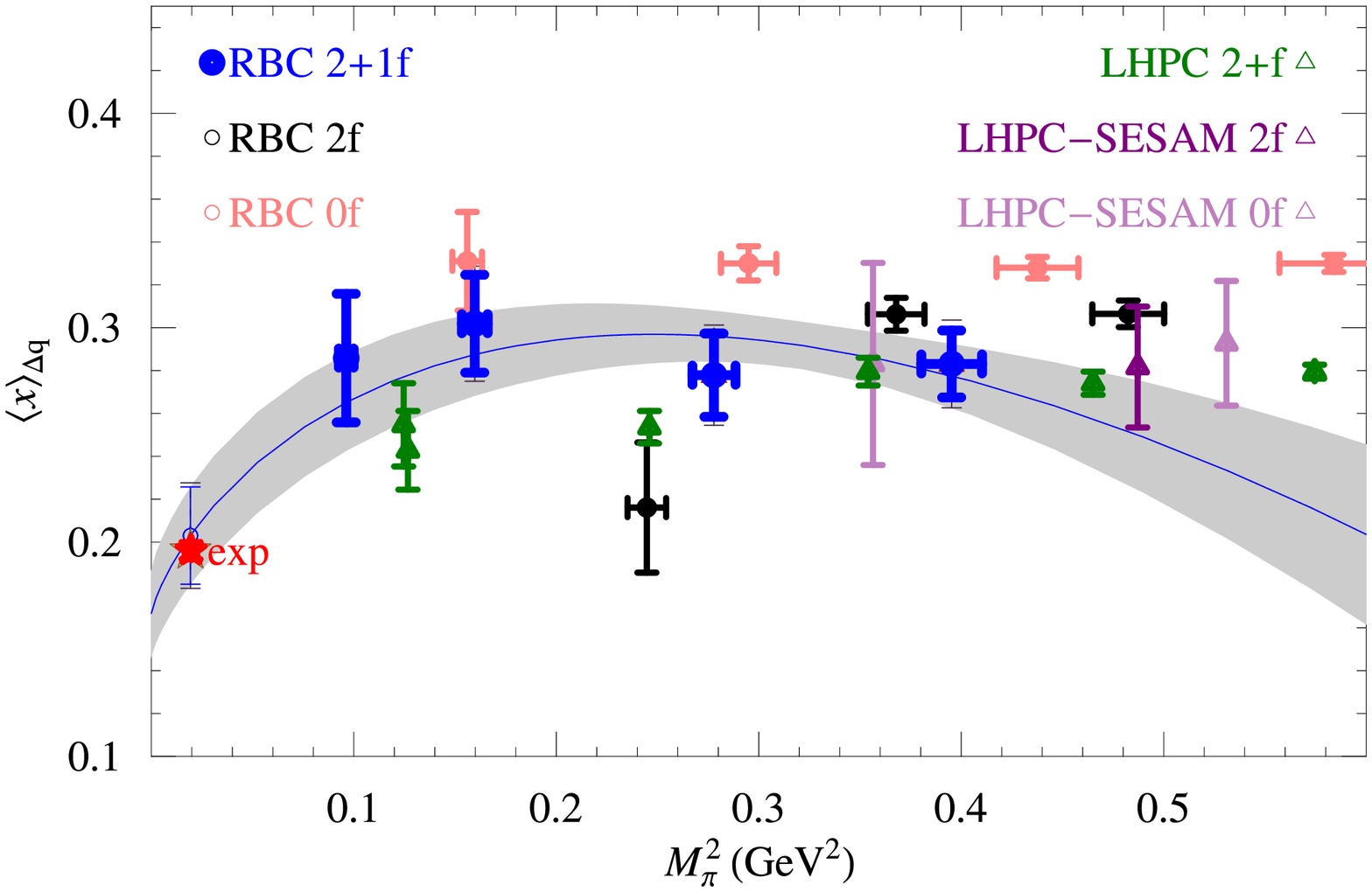}
\vspace{-0cm} \caption{Global comparison of the first moments of the unpolarized (left column) and helicity (right column) distributions, in terms of $m_\pi^2$  and their chiral extrapolations} \label{fig:allMoments}
\end{figure}

In this work, we will only concentrate on the first moments of the
unpolarized and helicity distributions, $\langle x\rangle_{u-d}$ and $\langle x\rangle_{\Delta u-d}$ respectively. The corresponding
renormalization constants are obtained using RI/MOM-scheme NPR
as described in Sec.~\ref{subsec:parameters}. The renormalized moments are shown in Figure~\ref{fig:moments} at each pion mass point. Again, we use help from chiral effective theory\cite{Detmold:2001jb,Arndt:2001ye,Chen:2001gr,Detmold:2002nf}
for these quantities to extrapolate to the physical pion point:
\begin{eqnarray}
\langle x\rangle_{u-d} &=& C \left[ 1
 - \frac{3g_A^2 + 1}{(4\pi f_\pi)^2}m_\pi^2
 \ln\left(\frac{m_\pi^2}{\mu^2}\right)\right]+e(\mu^2)\frac{m_\pi^2}{(4\pi f_\pi)^2} \\
 \langle x\rangle_{\Delta u- \Delta d} &=& \tilde{C} \left[ 1 -
\frac{2g_A^2 + 1}{(4\pi f_\pi)^2} m_\pi^2
\ln\left(\frac{m_\pi^2}{\mu^2}\right)\right]
+ \tilde{e}(\mu^2)\frac{m_\pi^2}{(4\pi f_\pi)^2}.
\label{eq:moments}
\end{eqnarray}
This chiral behavior is indicated in the blue line in Figure~\ref{fig:moments}. %
We see a strong curvature due to the chiral form; more light-pion points should be taken to eliminate extrapolation uncertainties. In the past, we have been finding these quantities to be about 50\% larger than the experimental ones. (See Ref.~\cite{RBC_2f,Orginos:2005uy} for example.)
In this updated 2+1f DWF calculation, these moments are 0.133(13) and 0.203(23). Figure~\ref{fig:allMoments} shows a list of the latest calculations of the first moments of the unpolarized (left) and helicity (right) distributions. Here we can see the quenched or partially quenched approximation results either from our past DWF calculation~\cite{RBC_2f,Orginos:2005uy} or from the QCDSF/LHPC\cite{Dolgov:2002zm,Orginos:2006lat}. Our preliminary results seem to be consistent with the LHPC's mixed action calculation\cite{Edwards:2006qx}, although more statistics in the near future will help us to clarify what role the staggered sea plays in these quantities.

\subsubsection{Transversity}\label{subsubsec:Trans}

Another interesting quantity regarding the spin structure of nucleon is transversity. The moments of transversity are defined as
\begin{eqnarray}
\langle x^n\rangle_{\delta q} &=& \int_0^1 dx x^n[\delta q(x)-(-1)^n\delta \overline{q}(x)],
\end{eqnarray}
where $\delta q$ is the difference between the quarks with spin aligned and anti-aligned with the polarized target. On the lattice, this corresponds to the matrix elements of the operator
\begin{eqnarray}
{\cal O}_{\mu\mu_1...\mu_n}^{\sigma q} &=& i^{n-1} \overline{\psi} \gamma_5\sigma_{\mu \{\mu_1}\overleftrightarrow{D}_{\mu_2}\cdot \overleftrightarrow{D}_{\mu_n\}}\psi.
\end{eqnarray}
Again, we only calculate the isovector quantity to eliminate the contribution from the disconnected diagrams.

\begin{figure}
\vspace{-0.5cm}
\includegraphics[width=0.49\columnwidth]{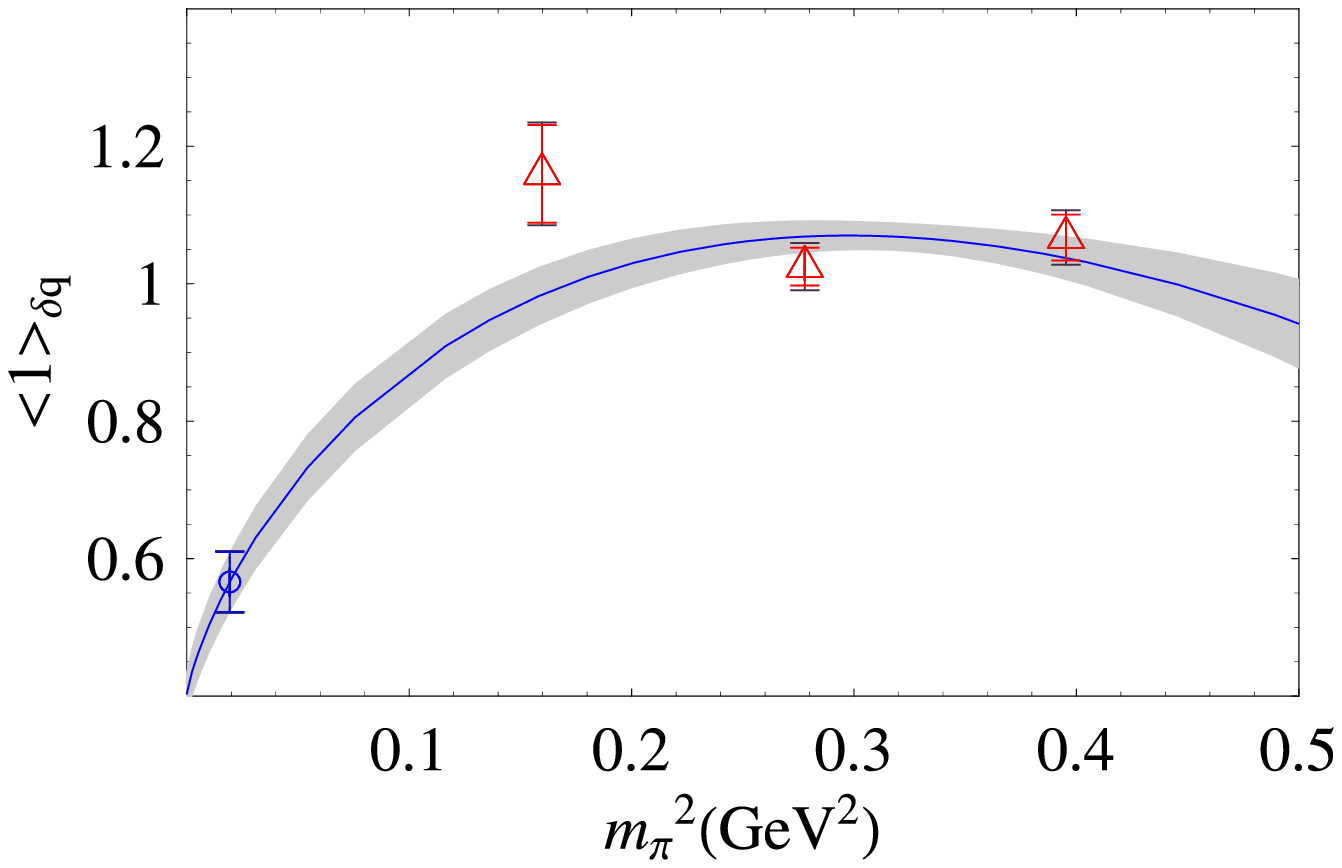}
\includegraphics[width=0.49\columnwidth]{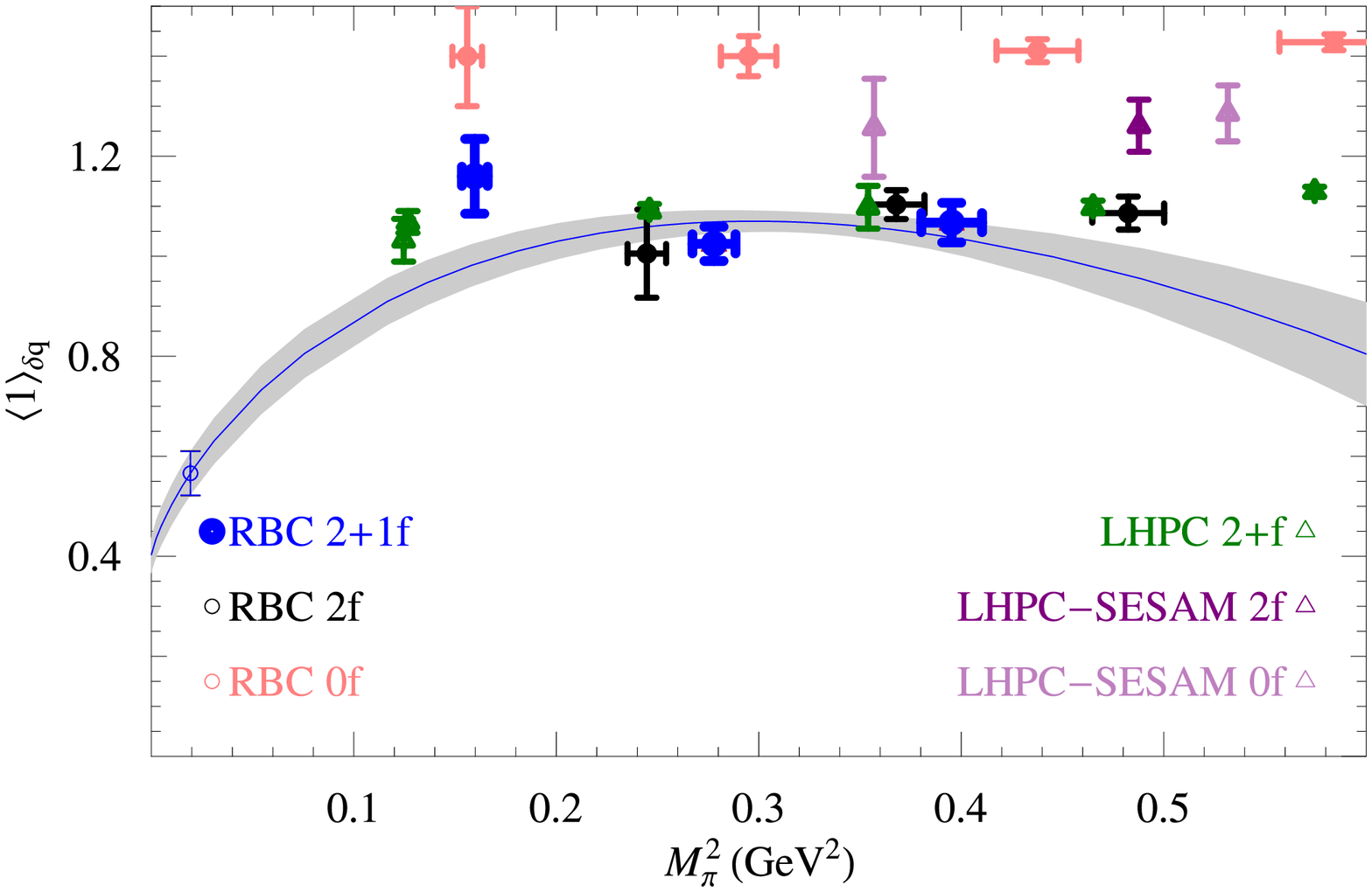}
\vspace{-0cm} \caption{Zeroth moment of transversity from our data: the chiral extrapolation to the physical pion point (left) and the global comparison among different lattice groups (right)} \label{fig:transversity}
\end{figure}

We calculate the zeroth moment of transversity, $\langle 1 \rangle_{\delta q}$, and the results are given on the left-hand side of Figure~\ref{fig:transversity}. We observe rather weak dependence (linear extrapolation) on the quark mass.  We use the chiral extrapolation formulation\cite{Detmold:2001jb,Arndt:2001ye,Chen:2001gr,Detmold:2002nf}
\begin{eqnarray}
 \langle x\rangle_{\delta u- \delta d} &=& \tilde{C}^\prime \left[ 1 -
\frac{4g_A^2 + 1}{2(4\pi f_\pi)^2} m_\pi^2
\ln\left(\frac{m_\pi^2}{\mu^2}\right)\right]
+ \tilde{e}^\prime(\mu^2)\frac{m_\pi^2}{(4\pi f_\pi)^2},
\label{eq:transversity}
\end{eqnarray}
and get 0.56(4) at physical pion mass point. This extrapolated value is significant smaller than the simulated pion mass point, which is of order 1 or so. We urgently need data from our lightest pion mass to confirm this rapid decreasing behavior. However, this is close to what has been found by LHPC with mixed action\cite{Edwards:2006qx}, about 0.7; their data is listed on the right-hand side of Figure~\ref{fig:transversity}. Their results at each pion mass are consistent within our current statistics. Apparently, the total suppression of sea quarks plays an important roles, as seen in comparing our quenched and dynamical numbers; however, there is not much difference between 2 and 2+1 flavors.

\subsubsection{$d_1$}\label{subsec:d1}

\begin{figure}
\vspace{-0.5cm}
\includegraphics[width=0.49\columnwidth]{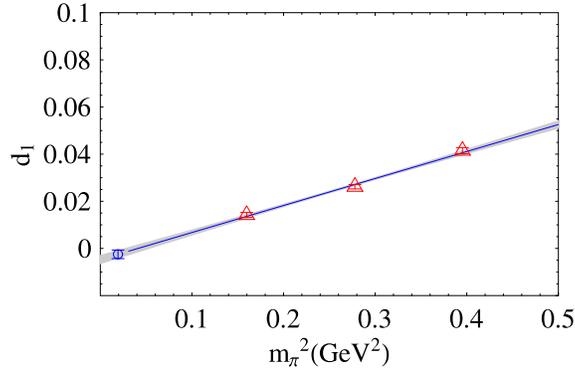}
\vspace{-0cm} \caption{Bare twist-3 matrix element $d_1$} \label{fig:d1}
\end{figure}

The twist-3 first moment of the polarized structure function $d_1$ is another interesting feature to consider. It is related to the polarized structure functions $g_1$ and $g_2$ by the operator
\begin{eqnarray}
{\cal O}^{5q}_{[34]} &=& i \overline{\psi} \gamma_5 [\gamma_3\overleftrightarrow{D}_{4} - \gamma_4\overleftrightarrow{D}_{3}]\psi.
\end{eqnarray}
It mixes with the lower-dimensional operator $O^{\sigma q}_{34}$ if the lattice fermions do not have chiral symmetry at finite lattice spacing; we are free of this problem in our calculation. Although it is not measurable in deep inelastic scattering of electrons on protons, it gives us some expectation of the higher moment $d_n$ matrix elements. Figure~\ref{fig:d1} shows our isovector $d_1$ matrix element results. We extrapolate the twist-3 matrix element to the physical pion mass and get $d_1^{\rm bare}=-0.002(2)$, which is consistent with zero. Combined with the small value of $d_2$ from QCDSF\cite{Gockeler:2000ja}, we conclude that the Wandzura-Wilczek relation between moments of $g_1$ and $g_2$\cite{Wandzura:1977qf}, which asserts vanishing $d_n$, is at least approximately true.

\section{Hyperons}\label{sec:hyperons}

\subsection{Lattice Parameters}
We use a mixed action, meaning that the sea and valence fermions use different discretizations. In our case, the sea fermions are 2+1 flavors of staggered fermions (in configuration ensembles generated by the MILC collaboration\cite{Bernard:2001av}), and the valence fermions are domain-wall fermions. The pion mass ranges from 360 to 700~MeV in a lattice box of size 2.6~fm. The strange-strange Goldstone is fixed at 763(2)~MeV, which unfortunately does not reproduce the physical strange-strange Goldstone mass. The gauge fields are hypercubic-smeared to improve the chiral symmetry of the fermion, and the fermion field is Gaussian-smeared to improve the signal. The source-sink separation is fixed at 12 time units. Table~\ref{tab:conf_Info} lists details of the configurations in use.

\begin{table}
\begin{center}
\begin{tabular}{c|ccccc}
\hline\hline
Label & $m_\pi$ (MeV) & $m_K$ (MeV) & $\Sigma$ conf. &  $\Xi$ conf.\\
\hline
m010  & 358(2)        & 605(2)      & 600            & 600\\
m020  & 503(2)        & 653(2)      & 420            & 436\\
m030  & 599(1)        & 688(2)      & 561            & 561\\
m040  & 689(2)        & 730(2)      & 306            & 319\\
\hline\hline
\end{tabular}
\end{center}
\caption{Configuration details}\label{tab:conf_Info}
\end{table}

\subsection{Strangeness Contribution}

Studying the strangeness content
of the nucleon is important to understanding QCD. Since the nucleon has zero net strangeness, any contribution to nucleon structure observables is a purely sea-quark effect. Many experiments are devoted to understanding strange quark contributions to the elctromagnetic form factors of the nucleon: HAPPEX\cite{Aniol:2005zf,Aniol:2005zg} and G0\cite{Armstrong:2005hs} at JLab, SAMPLE at MIT-BATES\cite{Spayde:2003nr}, and A4 at Mainz\cite{Maas:2004ta,Maas:2004dh}. The experimental results reveal a small but non-zero strange contribution to the proton elctromagnetic form factors.


To theoretically understand this nonperturbative physics, lattice QCD is a natural candidate for applying first principles. However, to extract individual quark contributions, the difficultly of calculating disconnected diagrams (as shown in Figure~\ref{fig:3pt}) must be taken into account. In the past, these diagrams have been directly calculated in the quenched approximation in lattice QCD\cite{Mathur:2000cf,Dong:1997xr,Lewis:2002ix}, giving values of $G_M^s$ ranging from $-0.28(10)$ to $+0.05(6)$. The Adelaide-JLab Collaboration used an indirect approach with the help of charge symmetry\cite{Leinweber:1995ie} and chiral perturbation theory to correct for quenching effects, obtaining $-0.046(19)$\cite{Leinweber:2004tc}. There does not seem to be consistency among these works. In this paper, we will use unquenched lattice data from mixed action with charge symmetry in the hope of bringing clarity to the chaos.

\begin{figure}
\includegraphics[width=0.49\textwidth]{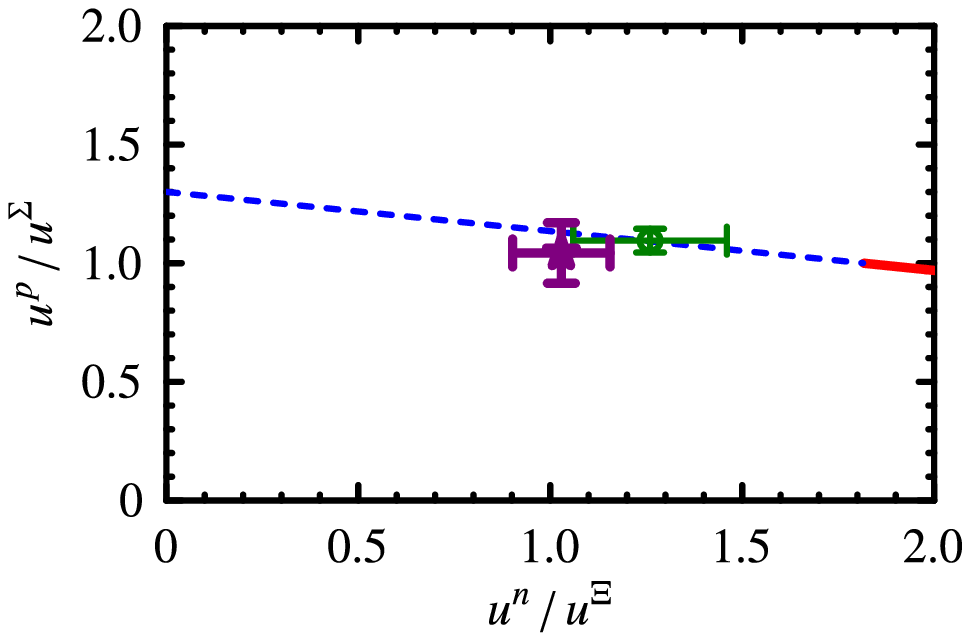}
\includegraphics[width=0.49\textwidth]{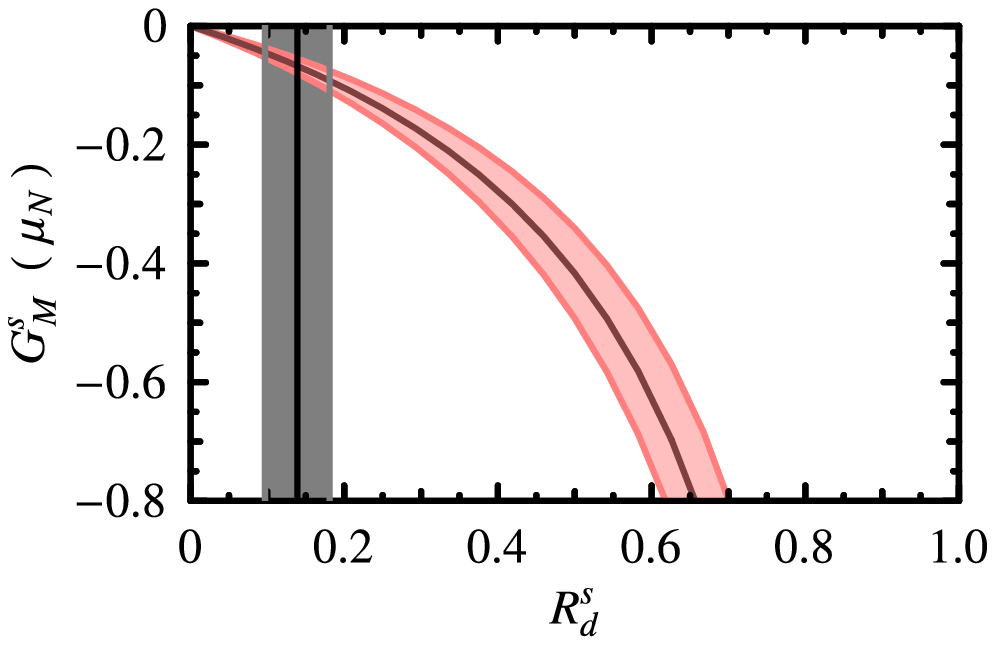}
\caption{
(Left) Our ${u^N}$ to ${u^\Sigma}$ and ${u^\Xi}$ ratio (purple star) compared with Adelaide-JLab Collaboration's (green circle). The blue dashed line indicates negative values of $G_M^s$, while the red solid line indicates positive ones. \\
(Right) The pink band is the constraint for the proton strangeness magnetic moment from our data and the grey band indicates the $R^s_d$ given from Ref~\cite{Leinweber:2004tc}.
}\label{fig:Adelaide}
\end{figure}

From charge symmetry\cite{Leinweber:1995ie,Leinweber:1999nf}, one can derive the following relation of magnetic moments ($\mu^B$) of individual quark contributions ($q^B$) and the disconnected contribution $O_B$ for each octet baryon $B$:
\begin{eqnarray} \label{eq:chargeSym}
\mu^p &=& e_u u^N + e_d d^N +O_N; \,;\;
    \mu^n = e_d u^N + e_u d^N +O_N;\nonumber \\
\mu^{\Sigma^+} &=& e_u u^\Sigma + e_s s^\Sigma +O_\Sigma; \,;\;
    \mu^{\Sigma^-} = e_d u^\Sigma + e_s s^\Sigma +O_\Sigma;\nonumber\\
\mu^{\Xi^0} &=& e_u u^\Xi + e_s s^\Xi +O_\Xi; \,;\;
    \mu^{\Xi^-} = e_d u^\Xi + e_s s^\Xi +O_\Xi. 
\end{eqnarray}
The disconnected contribution to the proton $O_N$ is
\begin{eqnarray}\label{eq:discoN}
O_N = \sum_q e^q G_M^q = \frac{1}{3}\frac{1-R^s_d}{R^s_d} G_M^s,
\end{eqnarray}
where $R^s_d$ is the ratio of the strange to down quark loop disconnected pieces, $G^s_M/G^d_M$. Combining Eq.~\ref{eq:chargeSym} and Eq.~\ref{eq:discoN}, we get the strangeness magnetic moment from either
\begin{eqnarray}\label{eq:GMs_sigma}
G_M^s = \frac{R^s_d}{1-R^s_d}\left[ 2\mu^p+\mu^n-\frac{u^N}{u^\Sigma}(\mu^{\Sigma^+}-\mu^{\Sigma^-})\right]
\end{eqnarray}
or
\begin{eqnarray}\label{eq:GMs_xi}
G_M^s = \frac{R^s_d}{1-R^s_d}\left[ \mu^p+2\mu^n-\frac{u^N}{u^\Xi}(\mu^{\Xi^0}-\mu^{\Sigma^-})\right].
\end{eqnarray}
The $\frac{u^N}{u^\Sigma}$ ($\frac{u^N}{u^\Xi}$) deviates slightly from 1 due to $SU(3)$ symmetry breaking. Using input from the well measured experimental quantities\cite{PDBook} $\mu^{\Sigma^+}-\mu^{\Sigma^-}=3.618$ or $\mu^{\Xi^0}-\mu^{\Sigma^-}=-0.599$, we get a constraint on the stangeness content of the proton magnetic moments.
The left of Figure~\ref{fig:Adelaide} shows the range of $G_M^s$: it goes to negative values as indicated by the blue dashed line, while the positive ones lie on the red solid line on the ratio parameter plane.
Note that since $(2\mu^p+\mu^n)-(\mu^{\Sigma^+}-\mu^{\Sigma^-})=0.057$, one would need higher precision on the lattice data for $\frac{u^N}{u^\Sigma}$ to make use of this constraint. Thus, in this work, we will only use Eq.~\ref{eq:GMs_xi} for $G_M^s$.

\begin{figure}
\includegraphics[width=0.49\textwidth]{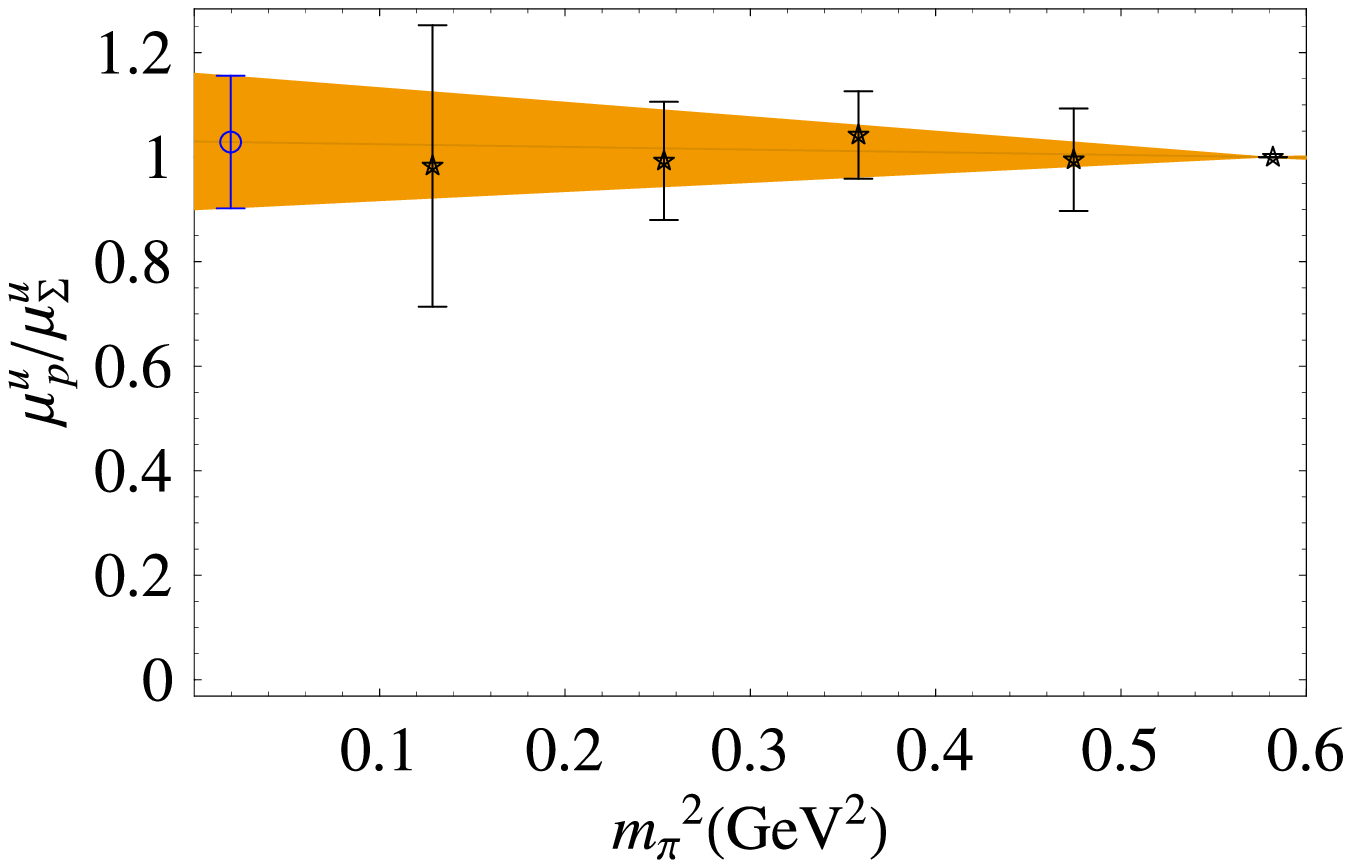}
\includegraphics[width=0.49\textwidth]{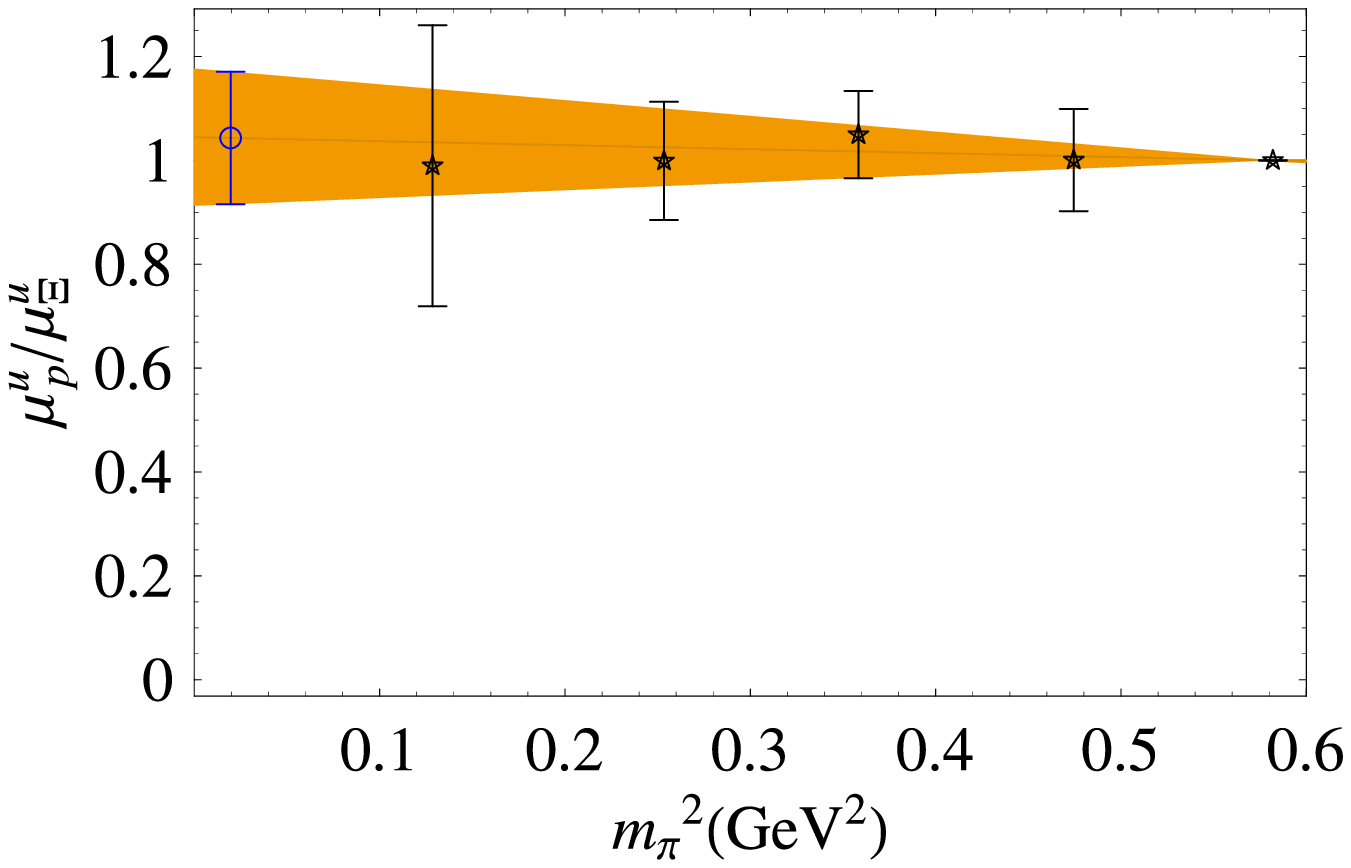}
\caption{Magnetic moment ratios of proton to $\Sigma$ and to $\Xi$ as functions of the pion mass}\label{fig:magneticRatio}
\end{figure}

We extrapolate to $q^2 = 0$ using a dipole form for the magnetic form factor of octet baryons. Figure~\ref{fig:magneticRatio} shows our full-QCD lattice data on the ratios $\frac{u^N}{u^\Sigma}$ (left) and $\frac{u^N}{u^\Xi}$ (right) as a function of $m_\pi^2$. The physical limit is taken by naive linear extrapolation, since the ratio might cancel out higher-order dependence on pion mass in the chiral extrapolation; the two ratios are 1.03(13) 
and 1.04(13) 
respectively.
We compare our extrapolated value with the calculation from Adelaide-JLab Collaboration\cite{Leinweber:2004tc} in the left part of Figure~\ref{fig:Adelaide}. We find our statistical errorbar in $\Sigma$ to be much larger than theirs. This might be because they use quenched configurations where the ensembles for different pion mass points are correlated; we extrapolated through different uncorrelated dynamical ensembles.
Combined with Eq.~\ref{eq:GMs_xi}, this gives a constraint (shown as a pink band in the right part of Figure~\ref{fig:Adelaide}) on the $G_M^s$ as a function of ratio of strange to up/down contribution to the proton $R^s_d$. Such a ratio will give smaller statistical error than calculating individual components, since fluctuations will be canceled in the ratio. However, at the moment, we do not have such a ratio calculated directly from the lattice. We quote the estimation $R^s_d=0.139(42)$\cite{Leinweber:2004tc} from chiral perturbation theory; this gives $G_M^s=-0.066(12)_{\rm stat}(23)_{R^s}$ with the dominate errorbar from the conservative estimation of $R^s_d$, which is consistent with Adelaide-JLab Collaboration's number, $-0.046(19)$.

\begin{figure}
\begin{tabular}{cc}
\includegraphics[width=0.49\textwidth]{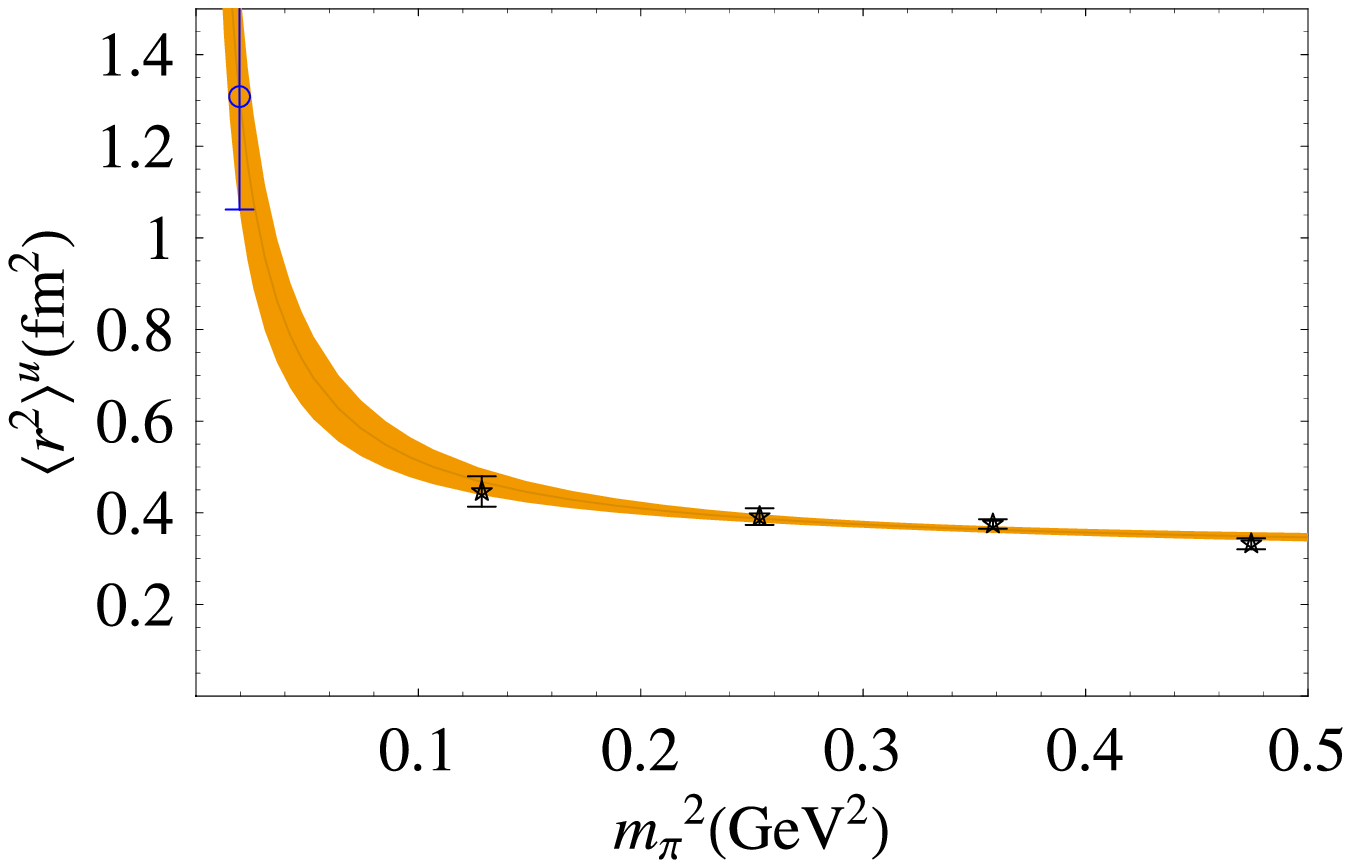}
&
\includegraphics[width=0.49\textwidth]{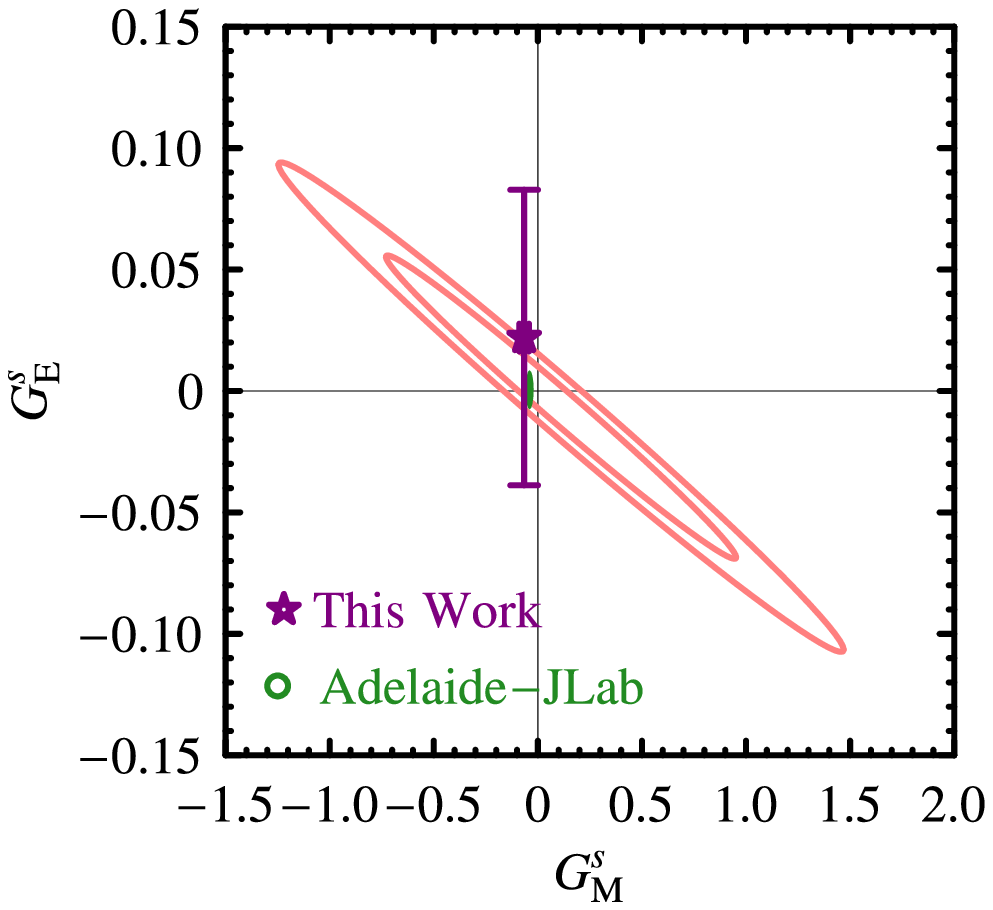}
\end{tabular}
\caption{
(Left) Our low statistics value of  $\langle r^2 \rangle^u$.
(Right) $G_M^s$-$G_E^s$ plane of the experimental region\cite{Young:2006mn} with our preliminary result from mixed action and Ref.~\cite{Leinweber:2006ug}
}\label{fig:GEGM}
\end{figure}

Similarly, with the help of charge symmetry, one can estimate the $G_E^s$ of the proton via the root-mean-square radius from the up quark connected diagram \cite{Leinweber:2006ug}:
\begin{eqnarray}
\langle r^2 \rangle^s = \frac{r^s_d}{1-r^s_d}\left[ 2\langle r^2 \rangle^p+\langle r^2 \rangle^n-\langle r^2 \rangle^u)\right].
\label{eq:GEs_u}
\end{eqnarray}
The left-hand side of Figure~\ref{fig:GEGM} shows the $\langle r^2 \rangle^u$ at each pion mass and their chiral extrapolation according to Ref.~\cite{Dunne:2001ip}. Note that this is the result from analyzing the number of configuration listed in Table~\ref{tab:conf_Info} only. Taking $r^s_d=0.16(4)$ from chiral perturbation estimates\cite{Leinweber:2006ug}, we find $G_E^s(Q^2=0.1\mbox{ GeV})=0.022(61)$, which is consistent with experimental values, as shown on the right of Figure~\ref{fig:GEGM}. The statistics will be further improved with the latest LHPC calculation.

\subsection{Hyperon Axial Coupling Constants}\label{subsec:hyperonAxial}

\begin{figure}
\includegraphics[width=0.49\textwidth]{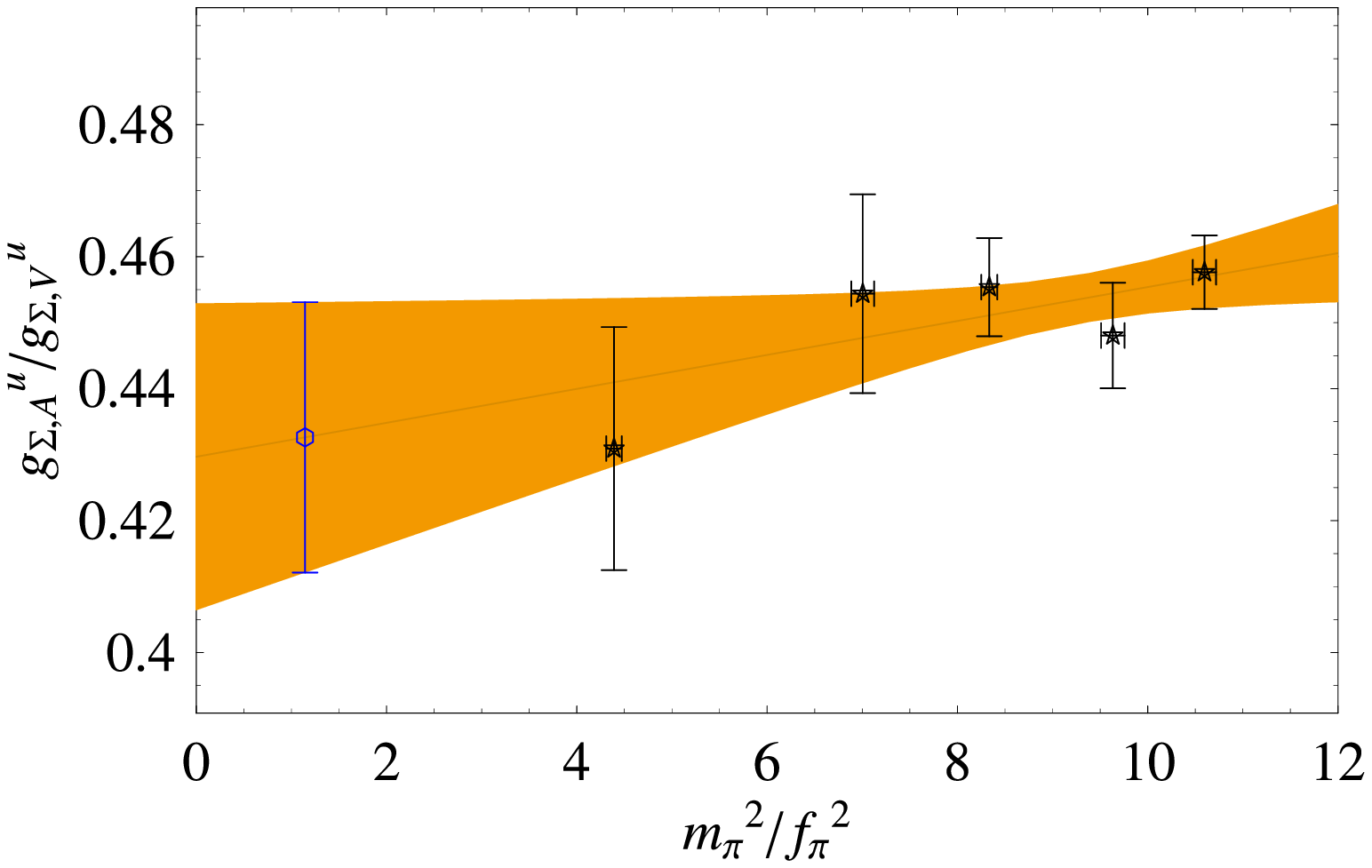}
\includegraphics[width=0.49\textwidth]{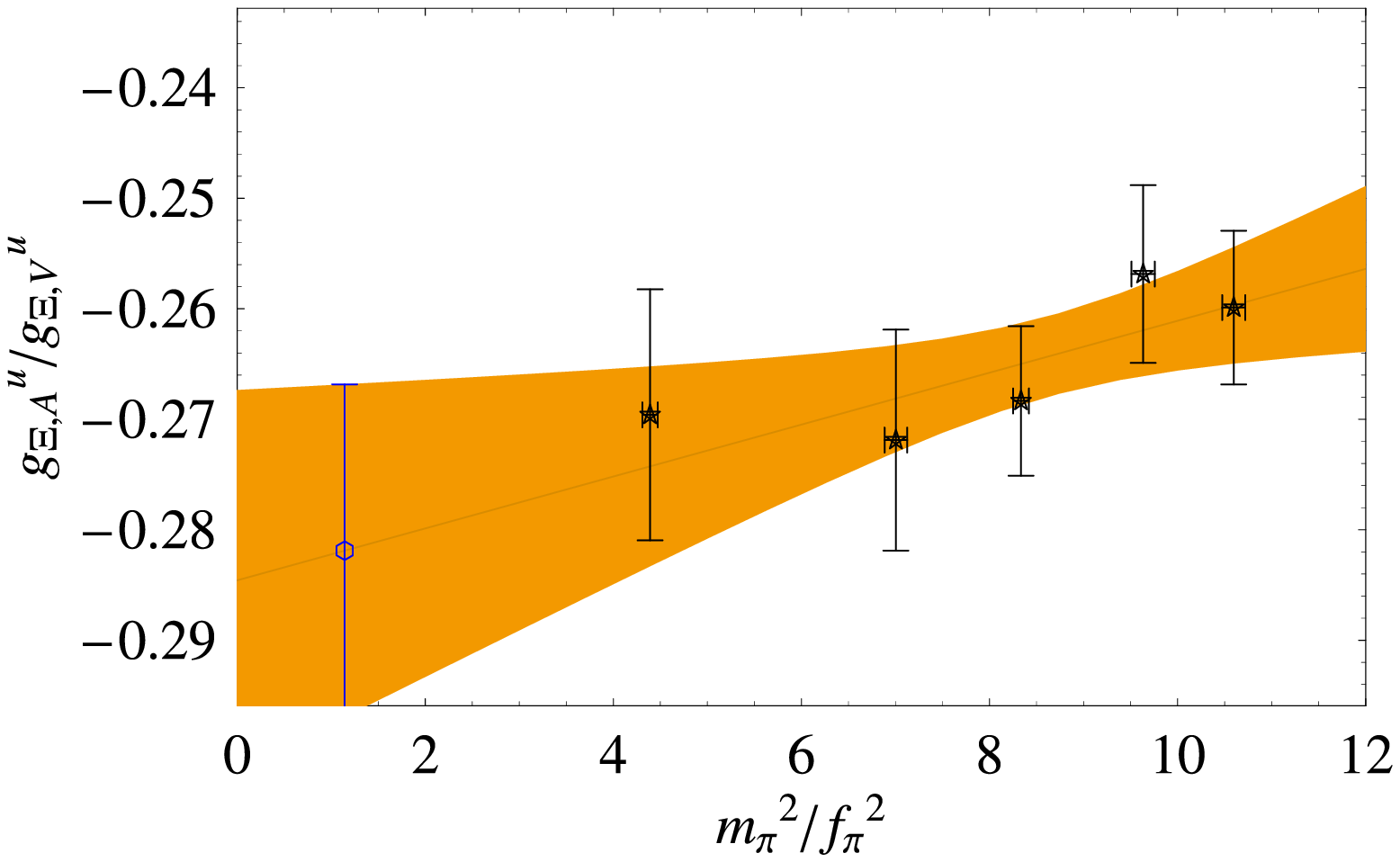}
\caption{Lattice data for $g_{\Sigma\Sigma}$ and $g_{\Xi\Xi}$ and chiral extrapolation
}\label{fig:hyperonAxial}
\end{figure}

The axial coupling constants, $g_{\Xi\Xi}$ and $g_{\Sigma\Sigma}$, have important applications such as in hyperon scattering and non-leptonic decays.
Previously, there only existed predictions from chiral perturbation theory\cite{Savage:1996zd} and large-$N_c$ calculations:
\begin{eqnarray}
&& 0.18 < -g_{\Xi\Xi} < 0.36\\
&& 0.30 < g_{\Sigma\Sigma} < 0.55.
\end{eqnarray}\label{eq:hyperonAxial}
Figure~\ref{fig:hyperonAxial} shows our lattice data and chiral extrapolation. Here we take a naive linear extrapolation against $(m_\pi/f_\pi)^2$. We find numbers consistent with the models, $g_{\Sigma\Sigma}= 0.441(14)$ and $g_{\Xi\Xi} = -0.277(11)$, but with much smaller errors.

\section{Conclusion and Outlook}
In summary, we report the latest nucleon calculations using RBC/UKQCD 2+1-flavor 3~fm DWF ensembles with pion masses as light as 310~MeV. Our calculations show good consistency with experimental values. Even those quantities, such as the first moments of the unpolarized quark distribution and helicity distribution, are chirally extrapolated to the physical pion mass in consistency with experimental values. We predict the zeroth moment of  transversity, and we find the twist-3 matrix element $d_1$ to be consistent with zero.

In analyzing hyperon form factors, we use an indirect approach to get the strangeness of the proton magnetic moment, using mixed action. We find our dynamical result to be consistent with Adelaide-JLab Collaboration's quenched result (which used a chiral correction for sea quark effects) and current experiments. The axial charge coupling of $\Sigma$ and $\Xi$ baryons are also predicted with significantly smaller errorbars than estimated in the past. We will continue to increase statistics, especially in the light pion mass region to get even more accurate results in the future.

\section*{Acknowledgements}
HWL thanks Riken-BNL center for their hospitality during work on the nucleon structure function and their support to make this trip; thanks also to collaborators Tom Blum, Shigemi Ohta, Kostas Orginos, Shoichi Sasaki and Takeshi Yamazaki for useful discussions; to Ross Young for discussion of strangeness physics and Anthony Thomas for helpful comments and discussions. The nucleon-structure calculations were done using the CPS on QCDOC with the resources of RBC collaboration, and the hyperon form factor calculations were performed using the Chroma software suite\cite{Edwards:2004sx} on clusters at Jefferson Laboratory using time awarded under the SciDAC Initiative. Authored by Jefferson Science Associates, LLC under U.S. DOE Contract No. DE-AC05-06OR23177. The U.S. Government retains a non-exclusive, paid-up, irrevocable, world-wide license to publish or reproduce this manuscript for U.S. Government purposes.

\bibliography{nuc_ref}

\end{document}